\documentclass[Preprint,12pt,3p]{elsarticle}

\usepackage{amssymb}
\usepackage{rotating}

\begin{document}

\begin{frontmatter}

\title{Nonmagnetic ground state, Crystal field effects and Heavy-fermion behaviour in the Remeika Phase: Pr$_3$Ir$_4$Ge$_{13}$}

\author[1,2]{K. Ramesh Kumar\corref{cor1} }
\author[2]{Michael O. Ogunbunmi \corref{cor2} }
\author[3]{Harikrishnan S. Nair}
\author[2]{Andr\'{e} M. Strydom}
\address[1]{Beijing National Laboratory for Condensed Matter Physics, Institute of Physics, Chinese Academy of Sciences,
Beijing 100190, China }
\address[2]{Highly Correlated Matter Research Group, Physics Department, University of Johannesburg, P.O Box 524, Auckland Park 2006, South Africa}
\address[3]{Department of Physics, 500 W. University Ave, University of Texas at El Paso, TX 79968, United States of America}

\
\cortext[cor1]{kraamesh57@gmail.com}
\cortext[cor2]{moogunbunmi@gmail.com, mogunbunmi@uj.ac.za}

\begin{abstract}
We report the magnetic, electronic and transport properties of the quasi-skutterudite compound Pr$_3$Ir$_4$Ge$_{13}$ by means of magnetic susceptibility $\chi(T)$, electrical resistivity $\rho(T)$, specific heat $C_p(T)$, thermal conductivity $\kappa(T)$, thermoelectric power $S(T)$ and Hall coefficient $R_\mathrm{H}(T)$ measurements.  Pr$_3$Ir$_4$Ge$_{13}$ does not show any phase transition down to 1.9 K. Magnetic, and specific measurements show that the system possesses a crystal electric field singlet ground state that is separated from the first excited state by about 37 K. $\rho(T)$ shows a negative temperature coefficient of resistance for the whole temperature range studied and which can be explained in terms of Mott's impurity band conduction mechanism. $R_\mathrm{H}(T)$ measurements show that Pr$_3$Ir$_4$Ge$_{13}$ is a low-carrier density semimetal and its transport properties indicate a metallic-non metallic cross over behaviour.  Large Seebeck values were observed for the entire temperature range of investigation, and the analysis of temperature variation of $S$ and $S/T$ showed no sign of strong correlation between the Pr 4$f^2$ and conduction electron states near Fermi level. A large Sommerfeld coefficient, $\gamma = 150$~mJ/(mol K$^2$) indicates the formation of a moderate heavy-fermion state emerging from the dynamical crystal field fluctuations.

\end{abstract}


\end{frontmatter}

\section{Introduction}
\label{intro}
Remeika and coworkers in 1980 have discovered a new family of ternary intermetallic compounds generally known as either Remeika phases or quasi skutterudites \cite{remeika1980new}. These compounds are structurally related to the filled skutterudite cage compounds and in these structures, a positively charged filler atoms are located inside a large size cage formed by a network of negatively charged covalent bonds\cite{gumeniuk2018structural}. Remeika phases are in general represented by the stoichiometry formula A$_3$T$_4$X$_{13}$  (A - Rare-earth/alkaline earth/actinide metal; T-Transition metal atoms; X = Ge, In and Sn)\cite{gumeniuk2018structural,Rai}. In this structure, six A and eight T atoms occupy the 6d (1/4 1/2 0) and 8$e$ (1/4 1/4 1/4) Wyckoff positions, respectively, and 26 X atoms occupy the 2$a$ (0 0 0) and 24$k$ (0 y z)  positions. The electronic character of the two X atoms is considerably different. The X' atom at 2$a$ site exhibits ionic character, whereas the X atom at 24$k$ shows a metallic bonding character, hence the general stoichiometry formula is seldom written as  X'A$_3$T$_4$X$_{12}$ instead of A$_3$T$_4$X$_{13}$\cite{gumeniuk2018structural}. More than 120 compounds have been reported in the literature that are exhibiting some exotic states of matter ranging from heavy-fermion state\cite{Hallas,yang2014kondo}, intermediate valence behaviour\cite{rai2016intermediate}, superconductivity\cite{strydom2014superconductivity,hayamizu2010superconducting}, charge density wave order\cite{wang2017electronic}, itinerant ferromagnetism\cite{gumeniuk2015magnetic}, and quantum criticality\cite{luo2018structural}. Among various classes, Tin containing compounds are extensively studied due to the interplay between structural, magnetic and superconducting properties. For example, (Ca, Sr)$_3$Ir$_4$Sn$_{13}$ exhibit charged density wave order at T = 33 K and 138 K respectively and the transition temperature can be suppressed by chemical or physical pressure to drive the system close to a quantum critical point \cite{Wang2012,luo2018structural}. Another interesting stannide based compound is Ce$_3$Co$_4$Sn$_{13}$ which shows complicated semiconducting property and exhibits single channel single ion impurity Kondo effect with $T_K$ (Kondo temperature) = 1.5 K \cite{cornelius2006observation,collave2015heavy}. Based on structural, magnetic, and electrical transport properties, it is observed that the Pr containing compounds crystallize in cubic structure with 3+ ionic state with no magnetic ordering and presence of crystalline electric field excitations. Pr$_3$Rh$_4$Ge$_{13}$ is a non-magnetic compound and exhibits moderate heavy-fermion behaviour ($\gamma$ = 100 mJ/mol K). Ramakrishnan et al. reported that the magnetic entropy release at 20 K is only 2 \% of R ln 2 ruling out the possibility of any long-range magnetic ordering \cite{ramakrishnan1996absence}. Low-temperature physical properties of Pr$_3$Co$_4$Sn$_{13}$ showed no magnetic ordering down to 0.2 K, and broad anomaly at 7 K in the magnetic heat capacity implies splitting of the  9-degenerate ($J=4$; $J$-total angular momentum number) levels due to the presence of crystalline electric field (CEF). Recently, Nair et al. have studied the ground state properties of Pr$_3$Rh$_4$Sn$_{13}$ through inelastic neutron scattering studies and low temperature heat capacity measurements\cite{nair2018absence}. The compound showed no magnetic ordering till 0.2 K, and the Schottky anomaly of the magnetic heat capacity is modelled using CEF scheme with seven energy levels. This energy level scheme is in close agreement with the excitation energy seen in the inelastic neutron scattering intensities\cite{nair2018absence}. 
Further, due to the formation of the cage network, the filler atoms are subjected to anharmonic vibrations known as \textquoteleft rattling\textquoteright  \cite{umeo2005probing,christensen2006crystal,nolas2000structural}. This effect suppresses the thermal conductivity of the systems due to phonon scattering without affecting the transport of the charge carriers. Large thermoelectric power $S(T)$, low lattice thermal conductivity k$_L$ and a semimetallic activation behaviour are a rare blend of properties that are seldom observed among thermoelectric materials needed to optimize the dimensionless figure of merit $ZT = S^2T\sigma/\kappa$  \cite{suekuni2007cage}. Recently, Ogunbunmi and Strydom reported on Pr$_3$Os$_4$Ge$_{13}$  \cite{ogunbunmi2020promising}, a heavy-fermion semimetal with a relatively low thermal conductivity (1.61~W K$^{-1}$ m$^{-1}$), enhanced thermoelectric power (32.85~$\mu$V K$^{-1}$)  and significant $ZT$ (= 0.03) values at room temperature. These features are comparable to those of several clathrates around the same temperature.\\
The synthesis and crystal structure of Pr$_3$Ir$_4$Ge$_{13}$ was first reported by Venturini $et al.$  \cite{venturini1985nouvelles} however there are no reports on the low temperature physical properties measurements and analysis so far to the best of our knowledge. Here we present the detailed investigations of the low temperature magnetic, heat capacity, electrical and thermal transport properties of the Remeika phase compound Pr$_3$Ir$_4$Ge$_{13}$.

\section{Experimental Methods}
\indent
Polycrystalline samples of Pr$_3$Ir$_4$Ge$_{13}$ were prepared by arc melting the stoichiometric amount of high-pure elements ($\geq$ 99.9  wt.\%)  using an Edmund B{\"u}hler arc melting furnace. The samples were melted several times on a water-cooled Cu plate under high pure argon atmosphere. The melted ingot was then wrapped in a tantalum foil and heat-treated at 900$^\circ$C for 14~days in an evacuated quartz tube to ensure homogeneity. Powder X-ray diffraction measurements were done using a Rigaku SmartLab diffractometer employing Cu-K$\alpha$ radiation. Full Rietveld refinements were carried out using the GSAS software \cite{toby2001expgui, larson1988generalized}.  We present a detailed description of the crystal structure of the title compound in Section~\ref{crystal chemistry}.

\indent
The electrical transport properties were measured by a conventional four-probe method with gold wire contacts made by spot welding equipment, and specific heat measurement was carried out using a quasi-adiabatic thermal relaxation method. The electrical transport and specific heat measurements were carried out using a Physical Property Measurement System (Quantum Design  Inc. San Diego) in the temperature range of 1.9~K and 300~K. The thermal conductivity, electrical resistivity and thermoelectric power were measured simultaneously on the same sample with the same set of thermal and electrical contacts. Magnetic susceptibility and magnetization studies were carried out on a Magnetic Property Measurement System (also by Quantum Design) in fields up to 7~T and in the temperature range between 1.8 and 300~K.

\begin{figure*}[!t]
	\centering
	\includegraphics[scale=0.5]{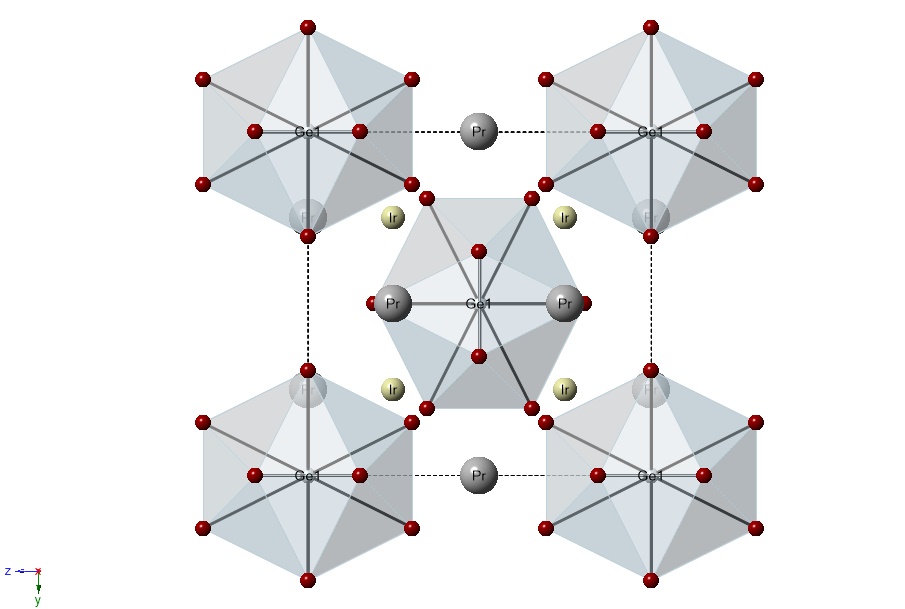}
	\caption{\label{Fig1Pm3n} (color online) Atomic arrangement of the Pr$_3$Ir$4$Ge$_{13}$ crystal structure projected along (100) direction. The polyhedral coordination shell represents the Ge1-Ge cage network formation. The grey and ivory colour balls are representing the Pr and Ir atoms, respectively. The Ge atoms are represented by the red spheres.}
\end{figure*} 
\begin{figure*}[!th]
	\centering
	\includegraphics[scale=0.5]{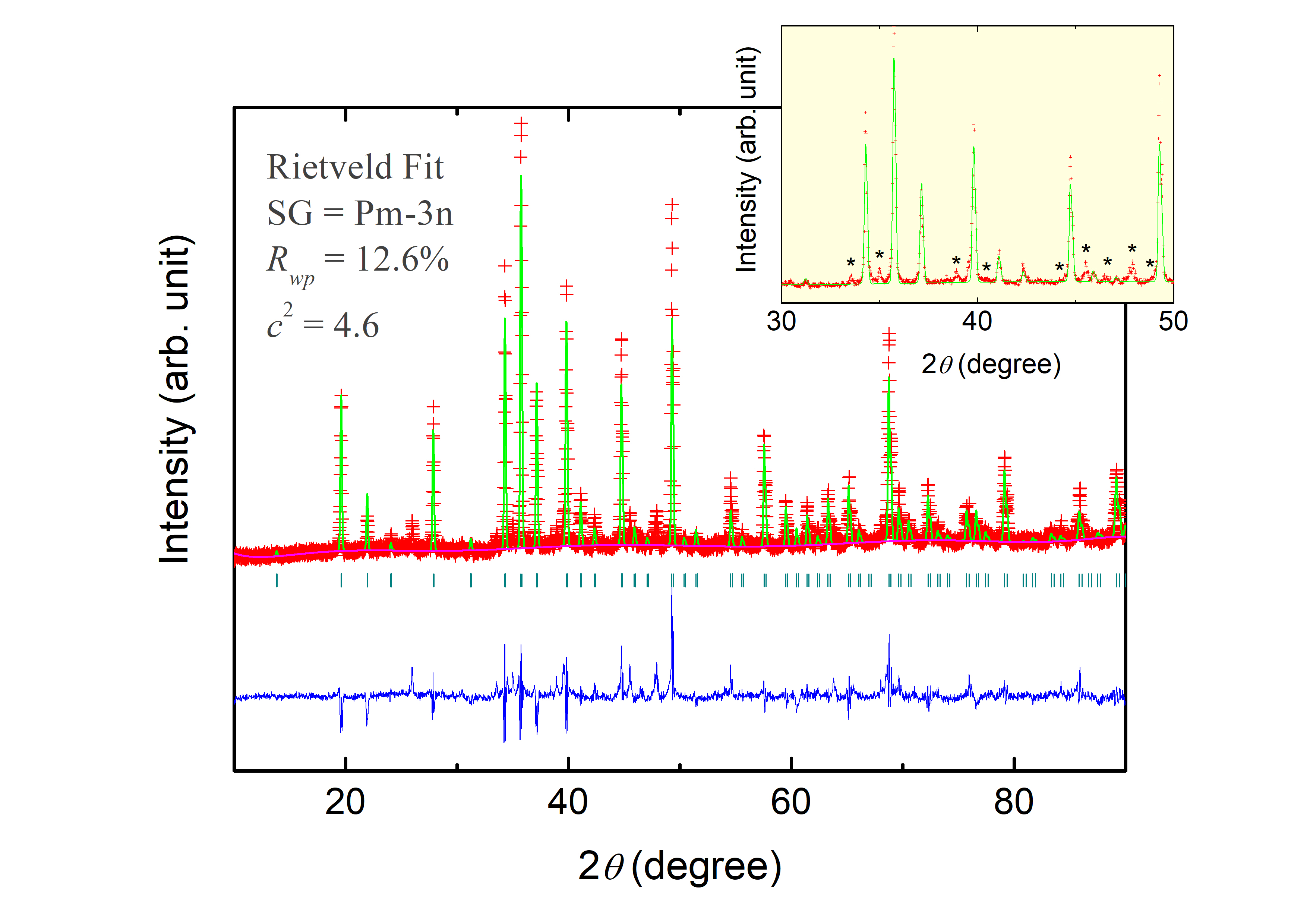}
	\caption{\label{Fig2_Rietveld} (color online) Experimental (red symbol) and calculated (green line) x-ray diffraction patterns of Pr$_3$Ir$_4$Ge$_{13}$. The blue vertical bars represent the allowed Bragg's reflections for the space group $Pm\overline3n$. The difference pattern is shown as black line at the bottom. The inset shows a portion of the x-ray pattern in a limited 2$\theta$ range and the asterisk symbols indicates the presence of superstructure reflections.}
\end{figure*} 
\section{Crystal chemistry and structure}
\label{crystal chemistry}
\begin{figure*}[!t]
	\centering
	\includegraphics[scale=0.5]{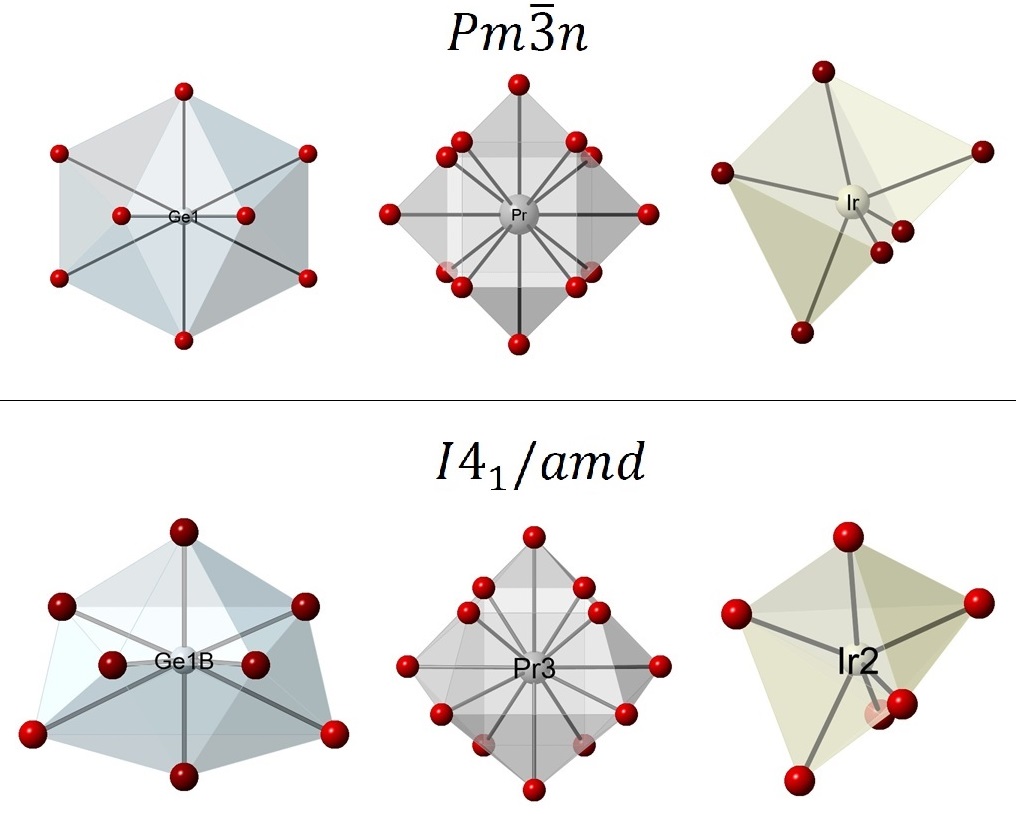}
	\caption{\label{Fig3_I4amd} (color online) Local site symmetry depictions of $Pm\overline3n$(top) and I4$_1$/amd (bottom) structural models: Ge'-(Ge$_{12}$) icosahedra (left), Pr-X$_{16}$ cuboctahedra (center) and transition metal trigonal prisms (right)}
\end{figure*}

\begin{figure*}[!h]
	\centering
	\includegraphics[scale=0.5]{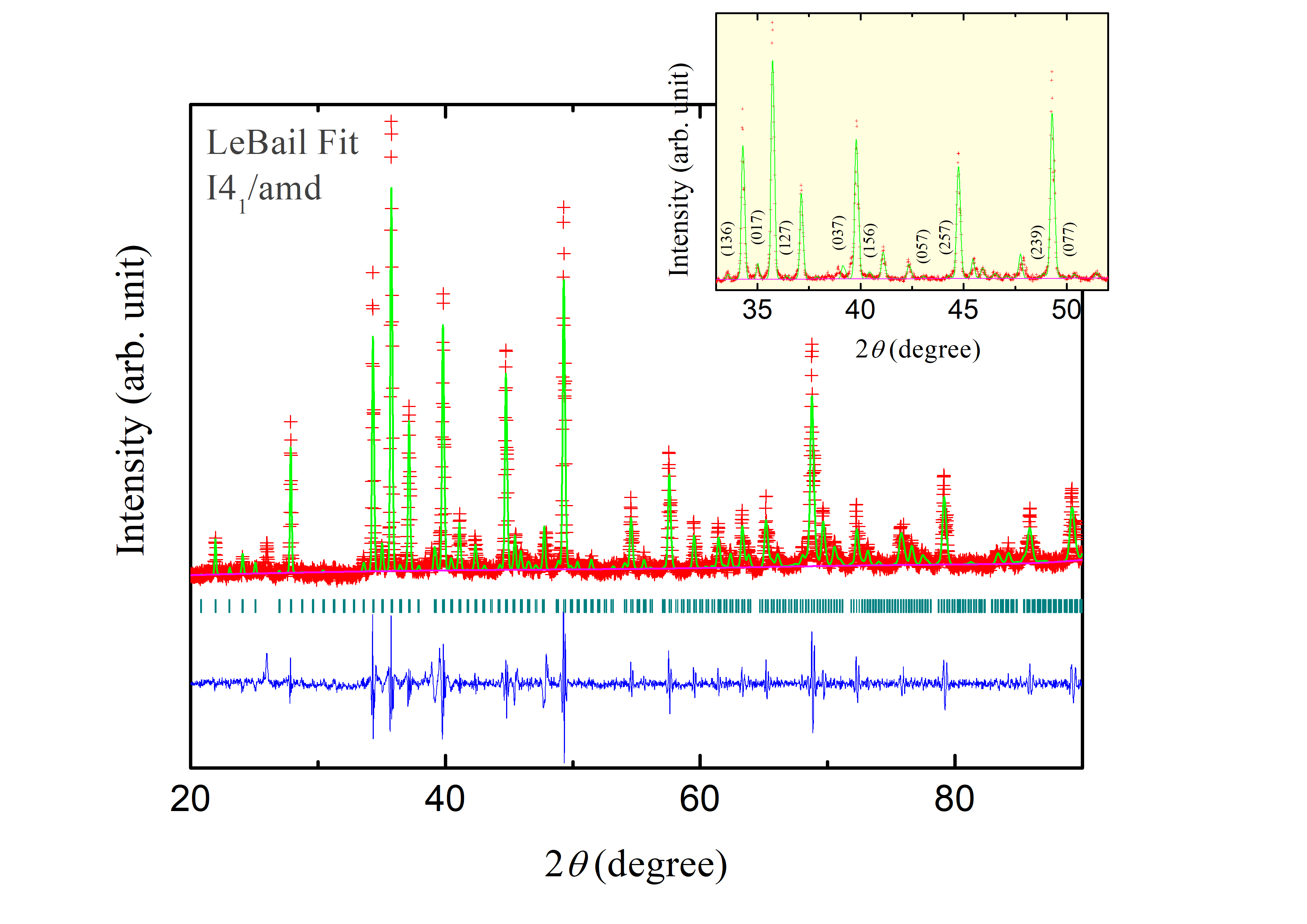}
	\caption{\label{Fig4_LeBail} 
		LeBail fitting for the powder XRD pattern of Pr$_3$Ir$4$Ge$_{13}$ by adopting I4$_1$/amd space group.  The red symbols and the solid continuous lines represent observed and calculated patterns, respectively. The difference plot is shown at the bottom. The inset shows the enlarged portion of the x-ray pattern and the allowed Bragg's reflections for I4$_1$/amd space groups are indicated by their corresponding Miller indices.}
\end{figure*}

The compound Pr$_3$Ir$_4$Ge$_{13}$ (Yb$_3$Rh$_4$Sn$_{13}$-structure type or Remeika phase) crystallizes in a cubic structure with space group  $Pm\overline3n$ (1).  Fig.~\ref{Fig1Pm3n} shows the atomic arrangement of Pr$_3$Ir$_4$Ge$_{13}$ crystal structure projected along (100) direction, and for the sake of clarity, we have shown only Ge'-(Ge$_{12}$)  icosahedron network. The experimental powder X-ray diffraction pattern along with calculated pattern ($Pm\overline3n$ structural model) are displayed in Fig 2.  The refined lattice parameter and atomic position values for the $Pm\overline3n$ structure model are shown in Table 1. The $Pm\overline3n$ structural model accounted only for the strong reflections, but did not account for several weak (superstructure) reflections (See Inset Picture of Fig.~\ref{Fig2_Rietveld}). Several authors have observed superstructure reflections in various Remeika phase compounds due to the presence of structural distortion and site disorders. Detailed analysis of the structural variants and group-subgroup relationship can be found elsewhere (1,3 and 4). Recently, Oswald et al. have observed a new structural variant belonging to a distorted tetragonal structure (space group I4$_1$/amd) in Lu$_3$Ir$_4$Ge$_{13}$  (2).  They observed that the Ge' site symmetry is lowered due to an elongation of the Ge'-(Ge$_{12}$) icosahedra along c-axis which in turn splits many of the major reflections corresponding to the cubic archetype (2). Fig.~\ref{Fig3_I4amd} shows the atomic arrangement and local site symmetry of A, T and X' atoms correspond to both $Pm\overline3n$ and I4$_1$/amd structural model. The Ge'-(Ge$_{12}$) icosahedra and Pr-X$_{16}$ cuboctahedra have shown considerable distortion in the tetragonal structure (Fig 3). Refining the low-resolution x-ray diffraction pattern by incorporating structural distortion/site disorder is a challenging task. Hence, we have employed LeBail fitting procedure to model the XRD pattern, which offers an easy and efficient way to test the plausibility of the available space groups. We used the atomic positions and occupancy values of Lu$_3$Ir$_4$Ge$_{13}$ as initial values for our fitting (2). Le Bail fit accounts for all the super-lattice reflections, and the reflections could be indexed using the distorted tetragonal structure I4$_1$/amd (Fig.~\ref{Fig4_LeBail}). The lattice parameter value was allowed to change during the refinement and the values were observed to be $a$ = 18.105 (1) {\AA} $b$ = 18.105 (1) {\AA} and $c$ = 18.065 (2) {\AA}.

\begin{table}[h!]
	\centering
	\caption{Crystal structure parameters obtained from the Rietveld
		refinement of the room-temperature powder x-ray diffraction data for Pr$_3$Ir$4$Ge$_{13}$. Standard deviations are given in the parentheses.}
	\begin{tabular}{c c c c c c c  }
		\hline
		\hline
		
		& Structure & Cubic \\
		& Space Group & $Pm\overline3n$  \\
		& a (\AA) &    \\
		
		\hline
		& Atoms & Wycoff position &  $x/a$ & $y/a$ & $z/a$  &\\
		\hline 
		&  Pr  & 6d  & 1/4 & 1/2 & 0  &    \\
		&  Ir  & 8e  & 1/4 & 1/4 & 1/4  &    \\
		& Ge'  & 2a  & 0 & 0 & 0 &     \\
		& Ge   & 24k & 0 & 0.142 (6) & 0.306 (6)  &     \\ \hline

	\end{tabular}%
	\label{tab:label}%
\end{table}%
\section{Magnetic properties}
\begin{figure}[!h]
	\centering
	\includegraphics[scale=0.5]{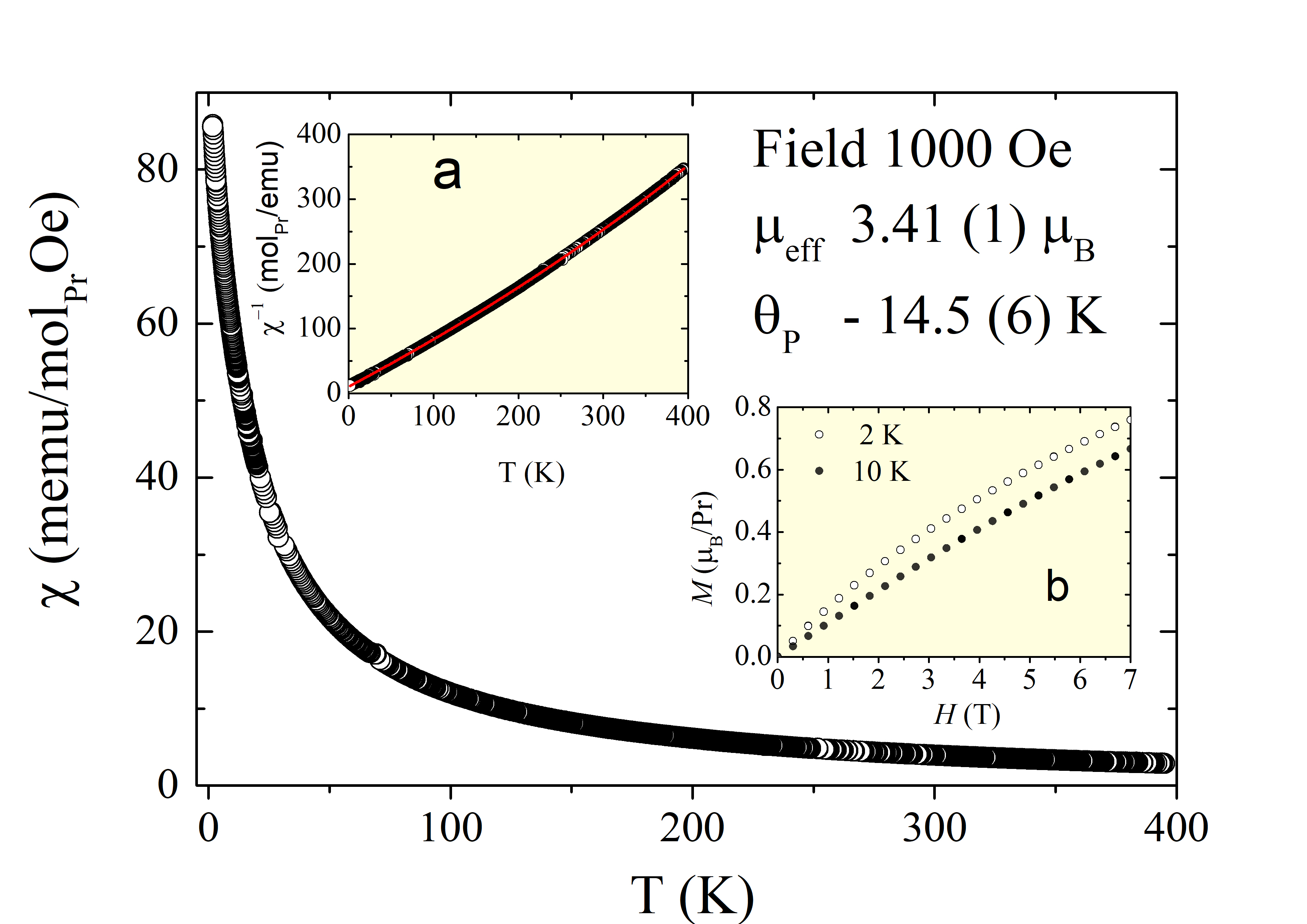}
	\caption{\label{Fig5_mag} Temperature dependence of magnetic susceptibility, $\chi(T)$ of Pr$_3$Ir$_4$Ge$_{13}$ measured in an external field of 0.05~T. Inset (a): Plot of $\chi^{-1}$ against temperature, $T$ and the red-solid line represents a linear fit using Curie-Weiss expression as described in the text. Inset (b): Isothermal magnetization at 2~K and 10~K.}
\end{figure}

The temperature dependence of magnetic susceptibility, $\chi(T)$ of Pr$_3$Ir$_4$Ge$_{13}$ is shown in Fig.~\ref{Fig5_mag}. No anomalies associated with magnetic order were observed throughout the temperature range of investigations. The temperature variation of the inverse susceptibility is modelled using the modified Curie-Weiss expression $\chi= \frac{C}{(T-\theta_p)}+\chi_0$ (Inset (a) of Fig.~\ref{Fig5_mag} ). In the above expression, $C$ and $\theta_\mathrm{p}$ stands for Curie-Weiss (CW) constant and paramagnetic Curie temperature, respectively. $\chi_0$ represents the temperature-independent part of the magnetic susceptibility, including the core-electron diamagnetism, Pauli paramagnetism and the second-order Van Vleck contribution. The relation between Curie constant and effective magnetic moment $\mu_\mathrm{eff}$ can be written as $C = N\mu_\mathrm{eff}^2$/3$k_\mathrm{B}$ where $N$ = total number of magnetic ions per formula unit. The least-squares fitting yields $\theta_\mathrm{P}$ = -14.6 (5) K, $\chi_0$ = -6.71 (2) $\times$ 10$^{-4}$ and $\mu_\mathrm{eff}$ = 3.41 (1) $\mu_\mathrm{B}$. The negative Curie-Weiss temperature indicates the magnetic correlation is antiferromagnetic and the estimated effective magnetic moment value is close to the expected value for Pr$^{3+}$ ions (3.53 $\mu_\mathrm{B}$). The isothermal magnetization data at 2 K and 10 K are shown in the inset (b) of Fig.~\ref{Fig5_mag}. The magnetization shows almost linear field dependence with no tendency of saturation till 7 T applied field. The magnetization value at 2 K for the 7 T field was observed to be 0.78 $\mu_\mathrm{B}$. The lack of magnetic order above 2 K and monotonic increase of $M$ with $H$ for Pr$_3$Ir$_4$Ge$_{13}$ indicate that the Pr$^{3+}$ ions have a nonmagnetic singlet ground state.

\section{Electrical transport}
\begin{figure}[!t]
	\centering
	\includegraphics[scale=0.6]{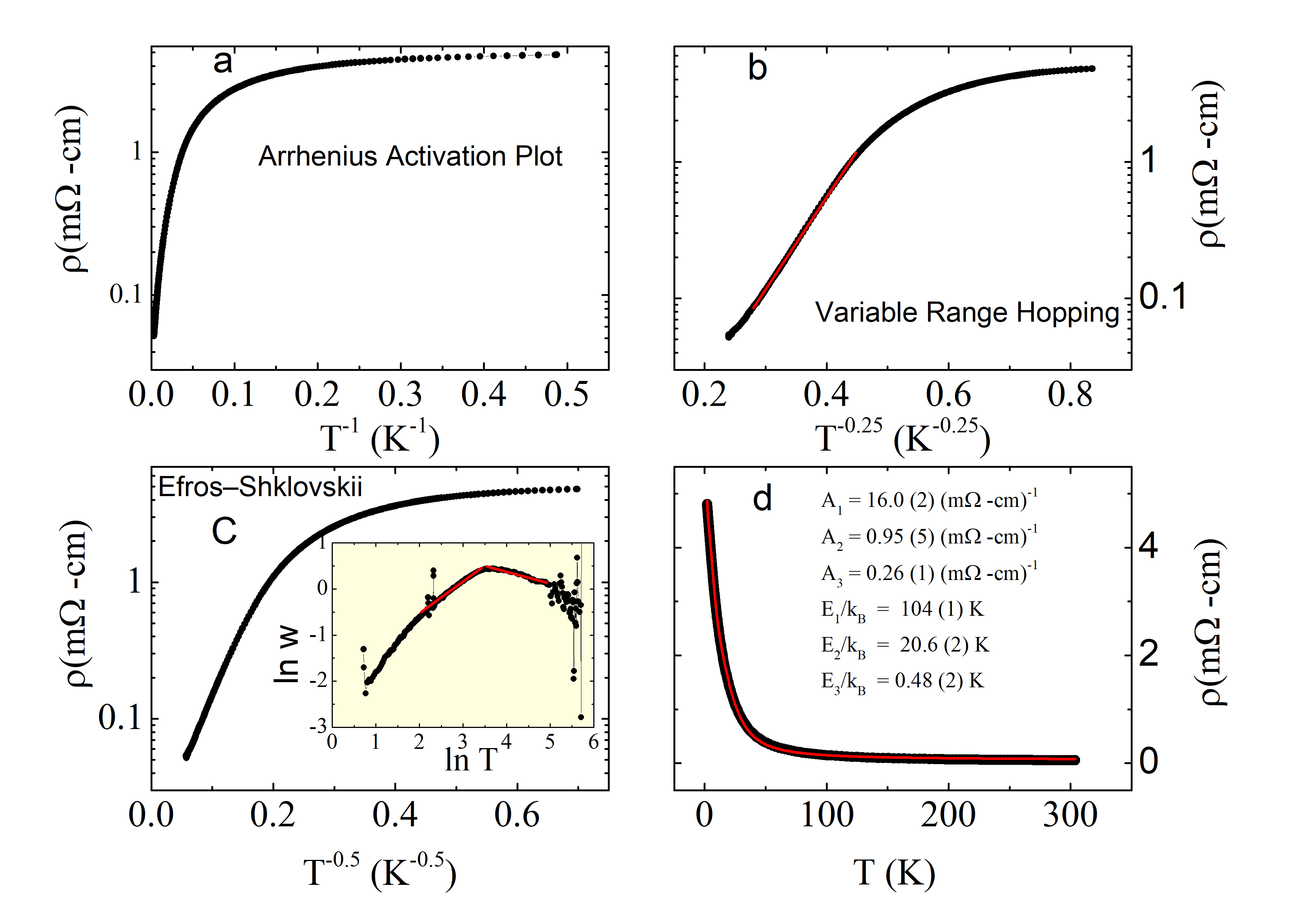}
	\caption{\label{Fig6_Resis} a) The $\rho$ as a function of $1/T$. (b) The resistivity plotted against $T^{−1/4}$, the red line is the Mott VRH fitting. (c) The resistivity plotted against $T^{−1/2}$ and the inset picture shows  $ln w$ variation as a function of $ln T$ along with a linear fit (red solid line) in two different regimes to emphasis the metal-nonmetallic cross over behaviour (d) The resistivity as a function of $T$, the red solid line represents the non linear fit based on Mott's impurity conduction mechanism.}
\end{figure}

Fig.~\ref{Fig6_Resis} (a-d) depicts the electrical resistivity as a function of various temperature exponents for the compound Pr$_3$Ir$_4$Ge$_{13}$ (a. ln $\rho$ vs $T^{-1}$, b. ln $\rho$ vs $T^{-1/4}$, c. ln $\rho$ vs $T^{-1/2}$, and d. $\rho$ vs $T$). The overall temperature variation of resistivity does not resemble a metallic behaviour as $d\rho/dT$ is negative for the whole temperature range of investigation.  Even though 3-4-13 germanidies show a negative temperature coefficient of $\rho$, the optical conductivity and the Hall effect measurements showed these compounds stay close to the border between metallic and nonmetallic state. Strydom \cite{Strydom} has observed a semi-metallic type activation behaviour and low charge carrier density in Y$_3$Ir$_4$Ge$_{13}$.  Rai et al., through magnetotransport, optical conductivity and thermodynamic measurements showed Yb$_3$Ir$_4$Ge$_{13}$ is a correlated semimetal and exhibits fragile magnetism \cite{Rai}. The temperature variation of $\rho$  and Hall coefficient ($R_H$) of  Pr$_3$Ir$_4$Ge$_{13}$ show similarity with the electrical transport property R$_3$Ir$_4$Ge$_{13}$ (R = Ce, Yb, Y and Lu) \cite{Strydom,Hallas,Rai}. Fig.~\ref{Fig6_Resis}a depicts the $\rho$ vs $1/T$ for Pr$_3$Ir$_4$Ge$_{13}$ and deviation from the Arrhenius activation behaviour is evident. To understand the conduction mechanism/cross over behaviour, we employed the following critical exponent analysis. The following universal formula can express the temperature variation of $\rho$.  
     \begin{equation}
\rho(T)=\rho_\alpha \mathrm{exp} \Bigg[\frac{T_\alpha}{T}\Bigg]^{n},
\label{Resis}
\end{equation}
     where $\rho_\alpha$, $T_\alpha$ and $n$ are resistivity parameter and characteristic temperature and critical exponent respectively \cite{Li}. For Mott's variable range hopping conduction the $n$ value is $1/4$ and for Efros-shkolvskii hopping conduction the $n$ value is $1/2$. Unlike Mott's variable range hopping condition, the Efros-shkolvskii hopping mechanism includes the long range coulombic interaction. Both these descriptions does not adequately explains the low temperature resistivity behaviour in Pr$_3$Ir$_4$Ge$_{13}$ (refer Fig.~\ref{Fig6_Resis}b-\ref{Fig6_Resis}c). Now by defining a quantity $w= \frac{-\partial(\mathrm ln \rho)}{\partial \mathrm ln T}$, equation 1 can be rewritten as 
     
 \begin{equation}
\mathrm {ln} W = \mathrm {ln} (nT_\alpha^n)-n\mathrm {ln} T
\label{Resis1}
\end{equation}
The above expression is a linear function with slope 'n' and if we plot $\ln W$ as a function of $\ln T$ change in the conduction mechanism, if any, can be observed in terms of the slope change  We adopt this analysis in our case to verify the metallic-nonmetallic cross over conduction mechanism. The inset of Fig 6d shows the variation of $\ln W$ as a function of $ln T$. We have observed a slope change at 32 K and our analysis indicates the sign of 'n' changes from positive to negative for decreasing temperature. If the conduction mechanism is governed only by normal activation behaviour and variable range hopping mechanism, then the $\ln W$-$\ln T$ plot 'n' value would be negative for whole temperature range, but our analysis showed a cross over behaviour. After ruling out the Mott's VRH mechanism, we attempted to model the resistivity variation by Mott's impurity conduction mechanism \cite{Mott}.  According to this theory, the temperature-dependent conductivity expressed as
\begin{equation}
\rho^{-1} = \sum_{i= 1}^{3} A_i \mathrm{exp} \frac{-E_i}{k_BT} \\
\label{Resis2}
\end{equation}

where $E_1$ is the activation energy for exciting an electron into the conduction band, $E_2$ is relevant in the compensated systems  and  $E_3$ is that for impurity conduction. Expression~\ref{Resis2} faithfully represents the experimental data with $A_1$ = 16.0 (5) (m$\Omega$-cm)$^{-1}$, $\frac{E_1}{k_B}$ = 104 (1) K, $A_2$ = .954 (2) (m$\Omega$-cm)$^{-1}$, $\frac{E_2}{k_B}$ = 20.9 (2) K, $A_3$ = 0.26 (1) (m$\Omega$-cm)$^{-1}$ and $\frac{E_3}{k_B}$ =  0.48 (2) K Fig.~\ref{Fig6_Resis}. A possible interpretation of this fit is that the large gap $\frac{E_1}{k_B}$ describes the intrinsic energy gap for Pr$_3$Ir$_4$Ge$_{13}$, whereas $\frac{E_2}{k_B}$ and $\frac{E_3}{k_B}$ describe impurity donor or acceptor states in the gap.
\\

\section{Hall Effect}
\begin{figure}[!t]
	\centering
	\includegraphics[scale=0.5]{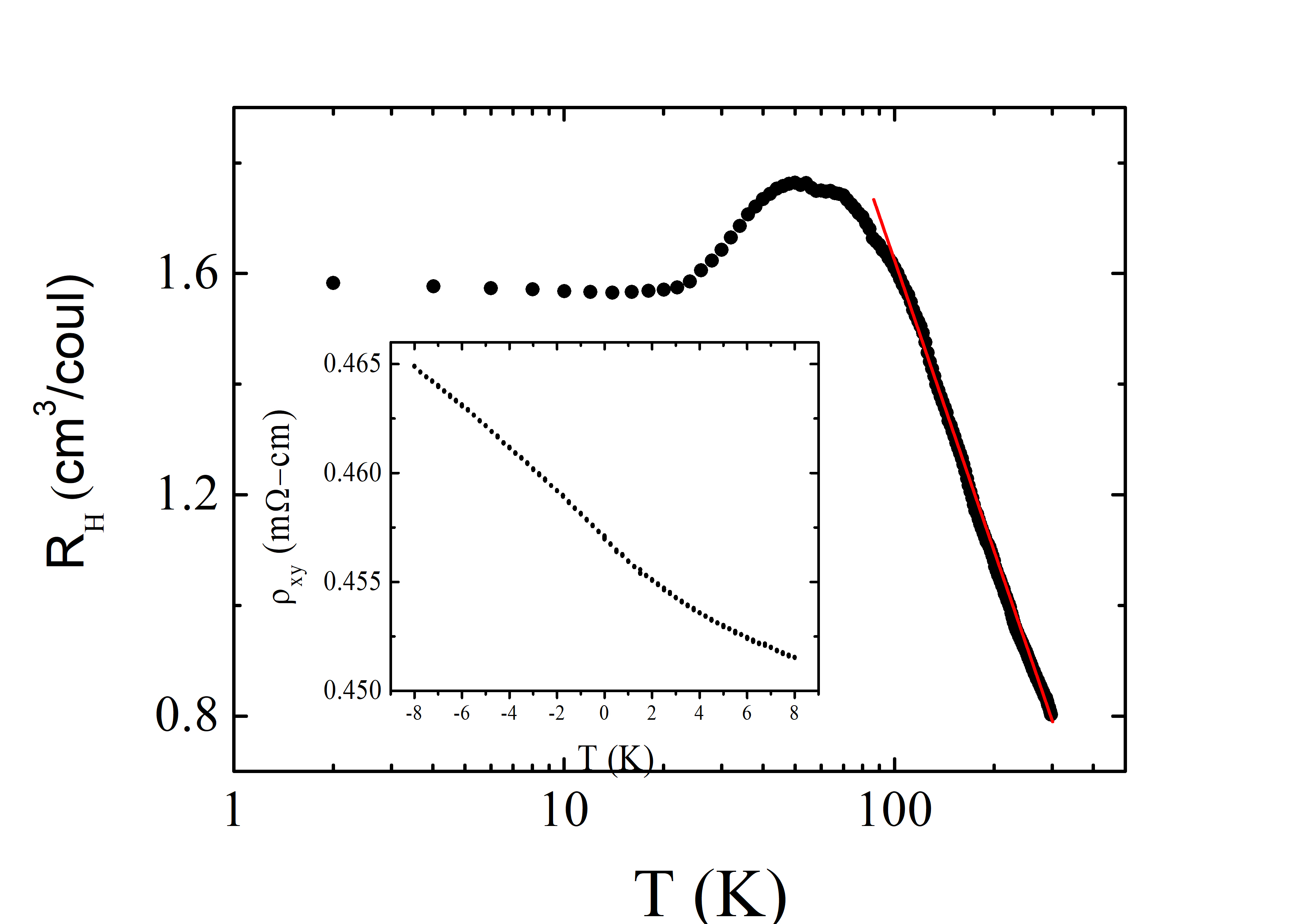}
	\caption{\label{Fig7_Hall} Temperature dependence of the Hall coefficient in a constant field of 6 T. The solid line on the data is a guide to the eye, and inset picture shows field variation of Hall resistivity at $T$ = 300 K.}
\end{figure}

We estimated the carrier density of Pr$_3$Ir$_4$Ge$_{13}$ through Hall effect measurement. Fig.~\ref{Fig7_Hall} shows the temperature dependence of the $R_H$ in the temperature between 1.9 K to 300 K in a semi-log plot. At high temperatures,  $R_H$ as a function of temperature resembles the temperature variation of $\rho(T)$, but at the low temperatures, the $R_H (T)$ deviates from the $\rho(T)$  and shows a plateau $\approx$ 50 K. The $R_H$ shows a drop in the same temperature range at which the $ln W$ vs $ln T$ exhibits a slope change. Y$_3$Ir$_4$Ge$_{13}$ has showed quite a similar temperature variation of $R_H$, but the reason for such plateau and low-temperature saturation of the $R_H$ are not given in detail \cite{Strydom}. Assuming single band model the Hall coefficient is related to the electron number density by the expression $R_H$ = -$\frac{1}{ne}$ where n is the electron number density and e is the electron charge. The carrier density below 30 K was estimated to 3.901 $\times$ 10$^{18}$ cm$^{-3}$ and the value is two orders of magnitude less than the value observed in Y$_3$Ir$_4$Ge$_{13}$. In an extrinsic semiconductor, as the impurity concentration N increases, the activation energy $E_3$ for the impurity conduction decreases and for a critical concentration, the $E_3$ vanishes. In such a scenario, $R_H$ becomes roughly independent of the temperatures at the low-temperature \cite{Mott}.  The estimated activation energy $E_3$ = 0.48 K corroborates  with our  $R_H$ analysis. Further, suppose the compensation ratio K (ratio between the number of acceptor levels to the donor levels) is large. then the charge carriers move in the sea of a random field arises due to positive (donor) and negative (acceptor) charges and such scenario does not facilitate to form a bound state or electron localisation \cite{Mott}. Hence the high conductivity, temperature-independent $R_H$ and metal to nonmetallic cross over behaviour in Pr$_3$Ir$_4$Ge$_{13}$ arise due to the transition between bound to free states for a single electron.
\section{Specific heat}
\begin{figure}[!t]
	\centering
	\includegraphics[scale=0.5]{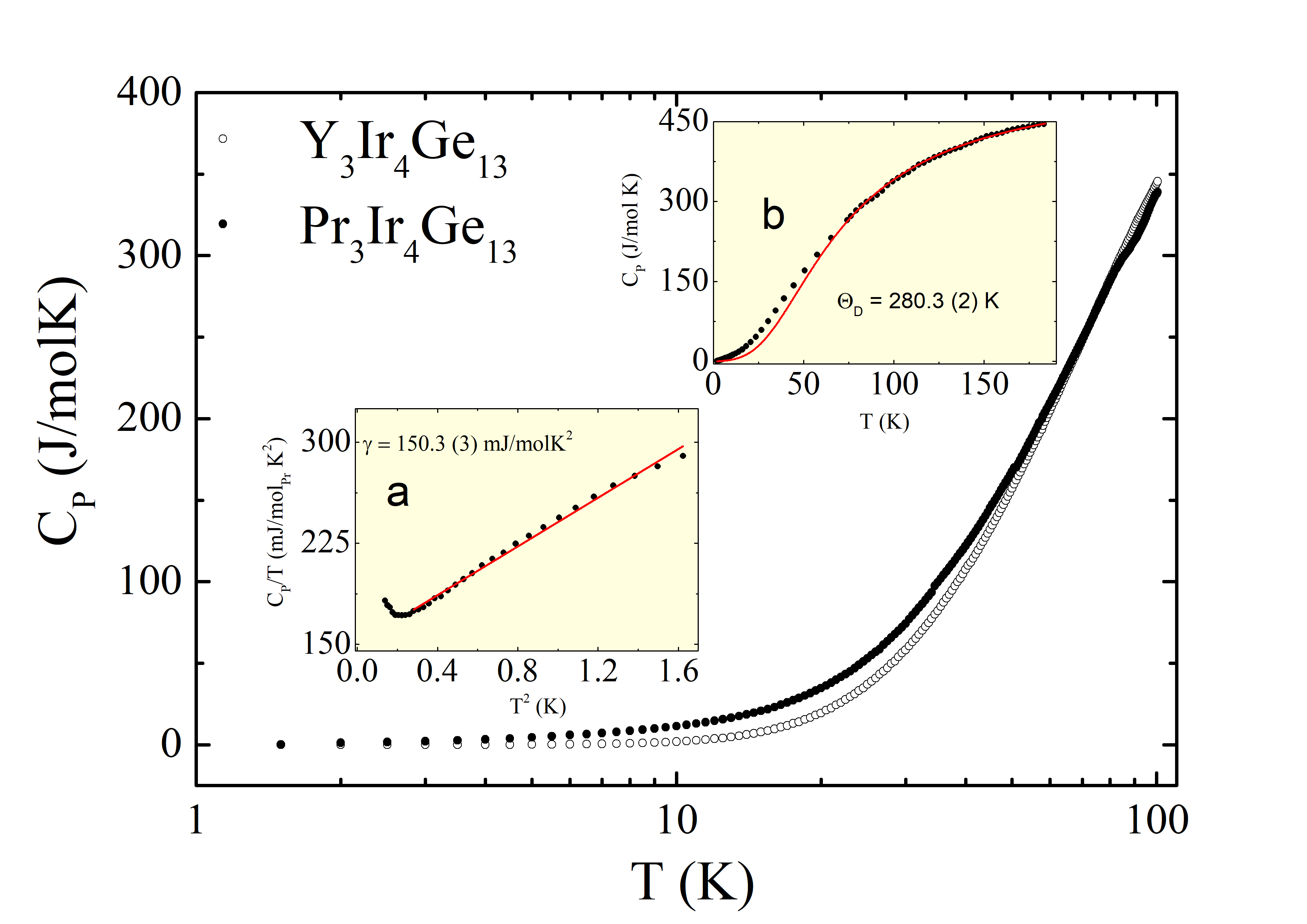}
	\caption{\label{Fig8_Cp} (a) Temperature dependence of specific heat, $C_p(T)$ of Pr$_3$Ir$_4$Ge$_{13}$ and Y$_3$Ir$_4$Ge$_{13}$ on a semi-$\log$ scale. Inset (a) presents $C_\mathrm{P}/T$ as a function of $T^2$ along with  linear fit (solid line). Inset (b) shows temperature dependence of $C_P$ along with nonlinear fit using Debye's heat capacity expression ~\ref{heatcapacity} }
\end{figure}

\begin{figure}[!t]
	\centering
	\includegraphics[scale=0.5]{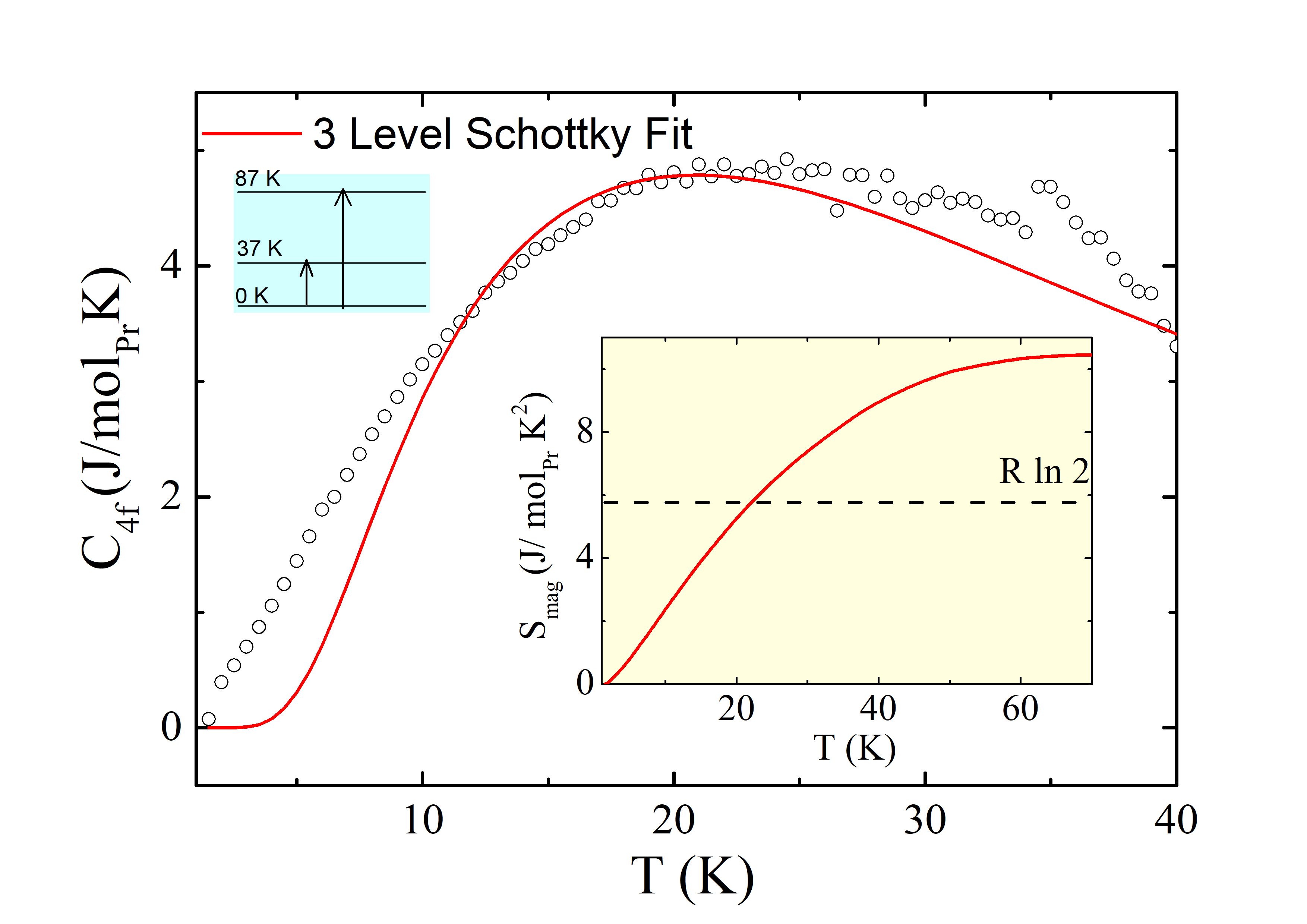}
	\caption{\label{Fig9_Smag} Magnetic specific heat as a function of temperature and the solid line represents the estimated magnetic heat capacity using the 3-level CEF model. The inset shows the temperature variation of magnetic entropy}
\end{figure}

The main panel of Fig.~\ref{Fig8_Cp} shows the molar specific heat of  Pr$_3$Ir$_4$Ge$_{13}$ and Y$_3$Ir$_4$Ge$_{13}$ on a semi-$\log$ scale in the temperature range between 2 K -- 300 K. In concert with magnetic susceptibility, the  $C_P$ vs $T$ did not show any long or short-range magnetic ordering above 1.9~K. The specific heat of a solid in the absence of low-frequency optical lattice vibration modes can be written as
\begin{equation}
C_P (T) = \gamma T+\Big[9NR(\frac{T}{\Theta_D})^3\int\limits_0^\frac{\Theta_D}{T}\frac{x^4}{(e^x-1)(1-e^{-x})}dx \Big] 
\label{heatcapacity}
\end{equation}
\noindent
The first and second terms in Eq. ~\ref{heatcapacity} represent the electronic specific heat and Debye lattice heat capacity, respectively.  $\Theta_D$ stands for Debye temperature, and $N$ represents the total number of atoms per formula unit. The inset (b) of Fig.~\ref{Fig8_Cp} shows the nonlinear fitting of the heat capacity using above expression, and it is clear from the analysis that below 100 K the phonon contribution is underestimated and the estimated Debye temperature is 280.3 (2) K. In the low temperatures, the heat capacity can be expressed as: 
\begin{equation}
C_P(T)/T = \gamma + \beta T^2; ~~ N(E_F)= \pi^2 k_\mathrm{B}^2 / 3\gamma,
\end{equation}
where $\gamma$ and $\beta$ are the electronic and phonon heat capacity coefficients, $N(E_F)$ is the density of states at the Fermi level and $k_\mathrm{B}$ is the Boltzmann constant. The $\gamma$ and $\beta$ values were estimated to be  150 mJ/mol K$^2$ (Inset (a) Fig. ~\ref{Fig8_Cp}). The same linear fitting has yielded  $\gamma$ = 4.3 (1) mJ/mol K  for Y$_3$Ir$_4$Ge$_{13}$  (\cite{strydom2007thermal}) indicating a moderate increase in the density of states near the Fermi Level for Pr$_3$Ir$_4$Ge$_{13}$.  The magnetic contribution $C_\mathrm{4f}$  to the heat capacity was estimated by subtracting the total heat capacity of Y$_3$Ir$_4$Ge$_{13}$ from the heat capacity of Pr$_3$Ir$_4$Ge$_{13}$. We scaled the heat capacity of Y$_3$Ir$_4$Ge$_{13}$ to account for the mass difference between these two materials by a term $P = \sqrt{\frac{M_{Pr}}{M_Y}}$ where $M_{Pr}$ and $M_Y$ are the molar mass of Pr$_3$Ir$_4$Ge$_{13}$ and Y$_3$Ir$_4$Ge$_{13}$ respectively. A broad Schottky-type anomaly centered around $\approx$ 20 K is observed in the magnetic part of specific heat  Fig.~\ref{Fig9_Smag}. The Schottky contribution to the specific heat arising from CEF effect is given by  expression
\begin{equation}
C_{Sch} = \frac{R}{T^2}\Bigg[\sum_{i=0}^{n-1}\frac{g_i\Delta_i^2e^{-\Delta_i/T}}{e^{-\Delta_i/T}}-\Bigg(\sum_{i=0}^{n-1}\frac{g_i\Delta_i^2e^{-\Delta_i/T}}{e^{-\Delta_i/T}}\Bigg)^2\Bigg] 
\end{equation}
where $g_i$s and $\Delta_i$s are degenracy and energy separation of corresponding CEF levels.

A CEF analysis with three levels reproduces the experimentally observed broad feature: 3 singlets at 0 K, 37 K and 87 K. As pointed out by Nair et al. through inelastic neutron studies and heat capacity measurements that the sketching the CEF energy level scheme of Pr containing Remaika phase compounds is a challenging task due to the presence of various structural distortions and site disorders. The more robust CEF level scheme warrants inelastic neutron scattering measurements to identify the excitation energies accurately. The magnetic entropy $S_{mag}(T)$ was calculated from the following expression 
\begin{equation}
S_{mag}(T) = \int\limits_0^T\frac{C_{4f}}{T'}dT'. 
\end{equation}

The ground state is a singlet which is separated by 37 K from the first excited is further supported from the fact that magnetic entropy attains a value of R ln 2 at 30 K (inset Fig.~\ref{Fig9_Smag})

. 
For typical HF systems, the ratio of $\chi(T)$ to $\gamma$ known as the Wilson ratio ($R_W$) \cite{wilson1975kg} is expected to give a value close to unity based on the expression:
\begin{equation}
R_W = \frac{\pi^2k_B^2 \chi (T\rightarrow 0)}{\mu_\mathrm{eff}^2 \gamma},
\label{SW_ratio}
\end{equation}

where $\mu_\mathrm{eff}$ is the effective magnetic moment,  $\chi (T\rightarrow 0)$ is the magnetic susceptibility in the low-temperature limit and all the other terms have their usual meaning. Using the observed low-temperature values of $\chi(T \rightarrow 0$) = 0.0856~emu/mol, $\gamma$ = 150~mJ/(mol$_{Pr}$ K$^2$) and $\mu_\mathrm{eff}$ = 3.41~$\mu_B$/Pr, a value of $R_W$ = 1.07 is obtained and which is comparable to a predicted value of unity.\\
\indent
\indent
\section{Thermal transport}
\begin{figure}[!h]
	\centering
	\includegraphics[scale=0.6]{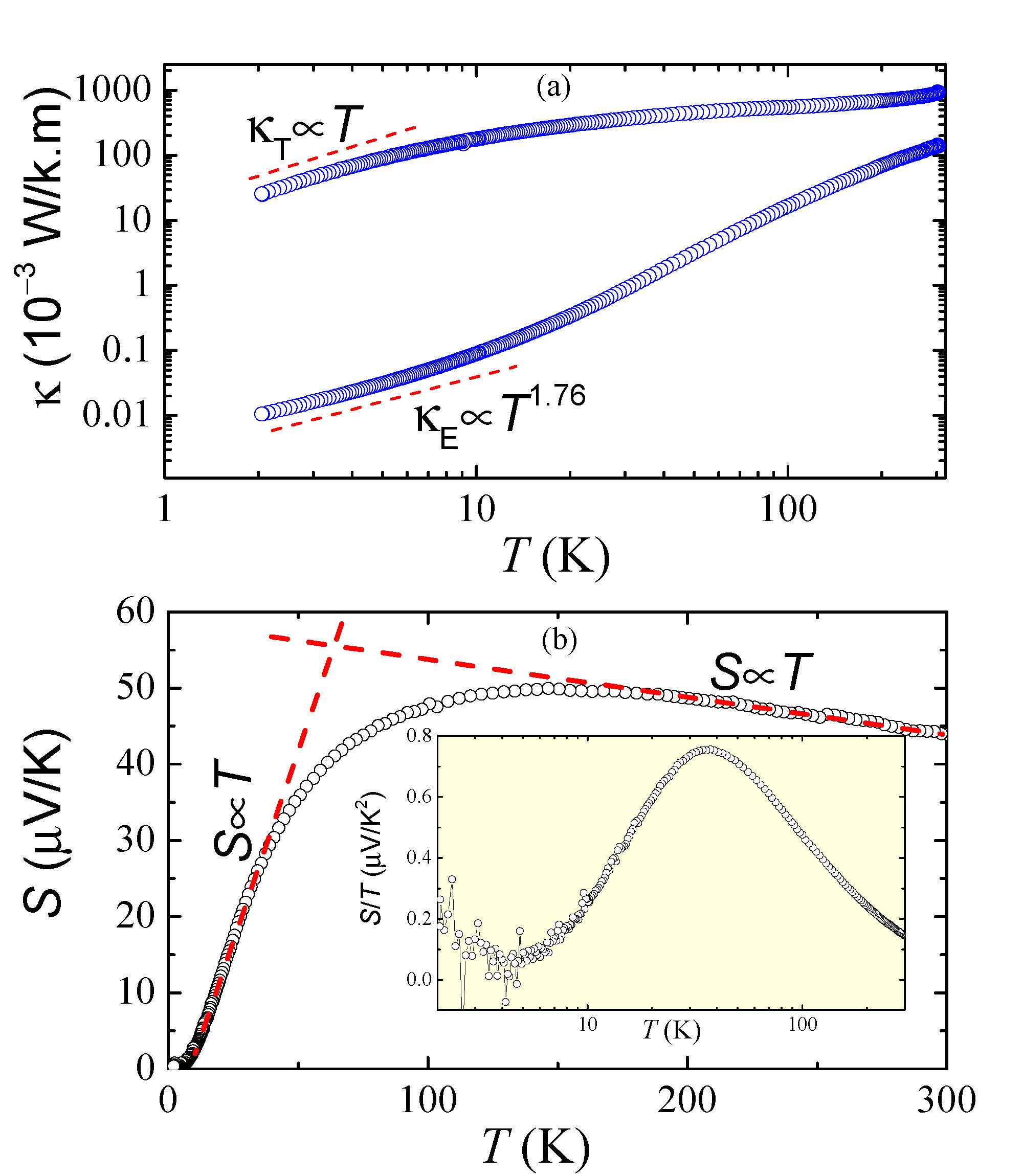}
	\caption{\label{fig_Pr3Ir4Ge13_tto} (a) Temperature dependence of total thermal conductivity, $\kappa_T(T)$ and electronic thermal conductivity, $\kappa_E(T)$ of Pr$_3$Ir$_4$Ge$_{13}$ on a $\log$-$\log$ scale. (b) Temperature dependence of the thermoelectric power $S(T)$ (main panel) and $S(T)/T$ (Inset)  for Pr$_3$Ir$_4$Ge$_{13}$. }
\end{figure}

To further characterize Pr$_3$Ir$_4$Ge$_{13}$, we measured the temperature dependence of thermal conductivity, $\kappa(T)$ and thermoelectric power, $S(T)$ simultaneously on a bar-shaped sample from room temperature down to 1.9~K. The temperature dependence of total and electronic thermal conductivities are shown in Fig.~\ref{fig_Pr3Ir4Ge13_tto} (a) on a $\log$-$\log$ axes. The overall $\kappa_T$ is very low, which is uncharacteristic of a good metal with a plateau-like behavior between 10 and 300~K. $\kappa_E$ is extracted from $\kappa_T$ using the Wiedemann-Franz relation \cite{kittel2005introduction}:
\begin{equation}
\kappa_E = L_0 T/\rho(T),
\label{eqn7}
\end{equation}
where $L_0$ is the Lorentz number which is given as $L_0$ = $\pi^2k_ B^2/3e^2$ = 2.45 $\times$ 10$^{-8}$~W$\Omega K^{-2}$, The electronic contribution $\kappa_E$ (Fig.~\ref{fig_Pr3Ir4Ge13_tto}) is 2-3 orders of magnitude less than $\kappa_T$ at low temperatures which implies negligible electronic contributions for the heat transport.  Hence the contribution due to $\kappa_E$ to the total thermal conductivity can be neglected, and the heat transport is predominately due to the lattice contributions to the thermal conductivity ($\kappa_T$ $\cong$ $\kappa_L$, where $\kappa_L$ is the lattice contribution). The dashed lines are guide to the eye showing the low temperature power law behaviour of $\kappa_T \propto T$ and $\kappa_E \propto T^{1.76}$ in the temperature range of 2~K $\le T \le$ 6~K and 2~K $\le T \le$ 10~K, respectively. Among, many factors that could be responsible for this observation are phonon-phonon scattering process and scattering from lattice defects in the polycrystalline material.

The total thermal conductivity value for Pr$_3$Ir$_4$Ge$_{13}$ ($\kappa_T$ $\simeq$ 0.8~W/Km) at room temperature is comparable to the value observed for U$_3$Ir$_4$Ge$_{13}$  (0.9~W/Km) \cite{gumeniuk2015magnetic} but slightly lower than a value ($\simeq$ 3~W/K.m) measured in the case of Y$_3$Ir$_4$Ge$_{13}$ \cite{strydom2007thermal}. 

\indent
The temperature variation of the thermoelectric power $S(T)$ is presented in  Fig.~\ref{fig_Pr3Ir4Ge13_tto}. Large positive Seebeck coefficient values in the entire temperature range of investigation imply that conducting mechanism is dominated by the holes type carriers near Fermi level. At room temperature, thermoelectric power value is observed to be 43.8~$\mu$V/K, and such high values indicate that the temperature variation of $S(T)$ is not governed by free-electron like a transport mechanism. In Heavy fermion systems, thermoelectric power is strongly influenced by the subtle changes in the electronic structure and the electronic heat capacity scales with the $S(T)$ in the low temperatures. In Pr$_3$Ir$_4$Ge$_{13}$, observation of large $\gamma$ and $S$ values would tempt us to draw an inference that Pr 4f$^2$ states are strongly correlating with conduction electron states at Fermi level. However, such a correlation is not valid in the present case as large $S$ values is observed even for the nonmagnetic compound Y$_3$Ir$_4$Ge$_{13}$ and hence the reason should be attributed to the unique electronic structure and semimetallic transport behaviour rather than renormalization of the 4f electronic states near the Fermi level. The overall behaviour of $S(T)$ can be modelled a linear expression in two different temperature regimes, separately. The $S(T)$ vs $T$ often exhibits complex behaviour due to different drag mechanisms, the influence of energy-dependent scattering rates and other quasi-particle excitations. Here, since $S(T)/T$ does not show any reasonable enhancement, heavy fermion state in Pr$_3$Ir$_4$Ge$_{13}$ emerges from the low lying crystal field fluctuations and exciton mediated electronic mass enhancement (See inset of Fig.~\ref{fig_Pr3Ir4Ge13_tto}).

\section{Conclusion}
In conclusion, we studied the structural, magnetic, electrical and thermal transport properties of the nonmagnetic heavy-fermion compound Pr$_3$Ir$_4$Ge$_{13}$.  Powder X-ray diffraction measurements and Rietveld analysis indicate that Pr$_3$Ir$_4$Ge$_{13}$ adopts a cubic structure (Yb$_3$Rh$_4$Sn$_{13}$-type structure) with space group $Pm\overline3n$ but presence of several unaccounted reflections point towards a structural distortion. LeBail fit using the $I4/amd$ space group accounts for all the super-lattice peaks which imply that $Pm\overline3n$ structure is only an average structure and R and Ge' atoms in this structure might adopt lower point symmetry (tetragonal distortion). The temperature dependence of dc susceptibility confirms the nonmagnetic nature of the sample above 2 K. The susceptibility follows Curie-Weiss behaviour in the high temperatures that imply localised nature of Pr$^{3+}$ moments. Overall temperature dependence of resistivity resembles a semimetallic type activation behaviour and $\rho(T)$ is modelled using Mott's impurity conduction phenomenon, including three energy gaps functions. Hall Coefficient measurements reveal that Pr$_3$Ir$_4$Ge$_{13}$ is low carrier density system with a clear metal-nonmetallic cross over behaviour at low temperatures. Specific heat analysis showed a significant enhancement of the linear heat capacity coefficient indicative of heavy-fermion behaviour. The thermal conductivity analysis shows that phonons largely dominate heat transport as the electronic contribution to thermal conductivity is negligible. Temperature variation of $S/T$ does not show an enhancement at the low temperature expected for strongly correlated electron systems; hence the observed heavy-fermion behaviour in Pr$_3$Ir$_4$Ge$_{13}$ stems from exciton mediated electronic mass enhancement.

\section{Acknowledgement}
KRK thanks the Institute of Physics, Chinese Academy of Sciences, Beijing for the International Young Scientist Fellowship. MOO acknowledges the UJ- GES 4.0 Post-doctoral fellowship.  AMS thanks the SA-NRF (93549) and UJ-URC for financial support.

\appendix

\bibliographystyle{elsarticle-num}

\bibliographystyle{elsarticle-num-names}

\end{document}